\documentclass[aps,prd,10pt,twocolumn,superscriptaddress,showpacs,footinbib]{revtex4-1}

\usepackage{hyperref}
\usepackage{graphicx}
\usepackage{amsfonts,amsmath,amssymb,bm,bbm}
\usepackage[english]{babel}
\usepackage{array}

\usepackage{dcolumn}

\newcolumntype{d}[1]{D{.}{.}{-1}}
\newcolumntype{C}[1]{>{\centering\let\newline\\\arraybackslash\hspace{0pt}}m{#1}}

\begin{document}

\title{First calculation of cosmic-ray muon spallation backgrounds for MeV astrophysical neutrino signals in Super-Kamiokande}

\author{Shirley Weishi Li}
\affiliation{Center for Cosmology and AstroParticle Physics (CCAPP), Ohio State University, Columbus, OH 43210}
\affiliation{Department of Physics, Ohio State University, Columbus, OH 43210}

\author{John F. Beacom}
\affiliation{Center for Cosmology and AstroParticle Physics (CCAPP), Ohio State University, Columbus, OH 43210}
\affiliation{Department of Physics, Ohio State University, Columbus, OH 43210}
\affiliation{Department of Astronomy, Ohio State University, Columbus, OH 43210 \\
{\tt li.1287@osu.edu,beacom.7@osu.edu} \smallskip}

\date{March 2, 2014}

\begin{abstract}
When muons travel through matter, their energy losses lead to nuclear breakup (``spallation'') processes.  The delayed decays of unstable daughter nuclei produced by cosmic-ray muons are important backgrounds for low-energy astrophysical neutrino experiments, e.g., those seeking to detect solar neutrino or diffuse supernova neutrino background (DSNB) signals.  Even though Super-Kamiokande has strong general cuts to reduce these spallation-induced backgrounds, the remaining rate before additional cuts for specific signals is much larger than the signal rates for kinetic energies of about 6 -- 18 MeV.  Surprisingly, there is no published calculation of the production and properties of these backgrounds in water, though there are such studies for scintillator.  Using the simulation code FLUKA and theoretical insights, we detail how muons lose energy in water, produce secondary particles, how and where these secondaries produce isotopes, and the properties of the backgrounds from their decays.  We reproduce Super-Kamiokande measurements of the total background to within a factor of 2, which is good given that the isotope yields vary by orders of magnitude and that some details of the experiment are unknown to us at this level.  Our results break aggregate data into component isotopes, reveal their separate production mechanisms, and preserve correlations between them.  We outline how to implement more effective background rejection techniques using this information.  Reducing backgrounds in solar and DSNB studies by even a factor of a few could help lead to important new discoveries.
\end{abstract}


\maketitle



\section{Introduction}\label{sec:intro}
Neutrinos are powerful probes of the universe and its contents.  They are abundantly produced by nuclear fusion processes that convert protons into neutrons, through the decays of unstable particles and nuclei created in high-energy processes, and through pair production in hot, dense environments.  They can reach us unattenuated and undeflected from vast distances or from behind enormous column densities of matter, directly revealing the energies and timescales of the processes that made them.  Even in a core-collapse supernova, where the neutrinos are thermalized by scattering, they emerge at energies $\sim 10$ MeV over about 10 s, compared to photons, which emerge at energies $\sim 1$ eV over months.  The  detection of astrophysical neutrinos allows us to probe physical conditions and neutrino properties beyond the reach of laboratory experiments.

The first great challenge of neutrino astronomy is the fact that the small interaction cross sections that make the above possible make detection difficult.  This can only be solved by brute force --- building large enough detectors to ensure adequate event rates.  We focus on Super-Kamiokande (Super-K), the world's largest low-energy neutrino detector, which has a fiducial mass of 22.5 kton of water and a total mass of 50 kton of water~\cite{Fukuda2003,Abe2014}.  (For comparison, neutrinos were first detected in the Reines-Cowan reactor experiment with a detector using less than 1 ton of scintillator~\cite{Cowan1956}.)  Even with such a large detector, the measured rates of low-energy astrophysical neutrinos are very small: about 15 solar neutrino events (all flavors of neutrinos elastically scattering electrons) detected per day~\cite{Hosaka2006,Cravens2008,Abe2011} and an upper limit of several events (primarily $\bar{\nu}_e$ inverse beta decay) detected per year from the diffuse supernova neutrino background (DSNB)~\cite{Malek2003,Horiuchi2009,Beacom2010,Bays2012,Zhang2013}.

The second and far greater challenge of neutrino astronomy is reducing detector backgrounds to isolate these rare signals.  Immense care and sophistication is required, and continual progress with existing detectors is possible.  The primary backgrounds for solar and DSNB signals are MeV electrons and positrons from the decays of nuclei and muons.  Below about 6 MeV detected electron kinetic energy, intrinsic radioactivities are the dominant background in Super-K~\cite{Gando2003,Hosaka2006,Cravens2008,Abe2011}, and these are controlled through the selection and purification of materials, choice of water circulation pattern to minimize radon ingress, and software processing (e.g., reconstruction quality and fiducial volume cuts).  From about 6 to 18 MeV kinetic energy, induced radioactivities produced by cosmic-ray muons are the dominant background~\cite{Gando2003,Hosaka2006,Cravens2008,Abe2011}, and there is great potential to reduce these with the help of theoretical work.  

To reduce cosmic-ray backgrounds, Super-K was built under 1000 m of rock (2700 m water equivalent) in the Kamioka mine in Japan~\cite{Fukuda2003,Abe2014}.  As cosmic-ray particles interact with the rock and lose energy, their flux is reduced. The only high-energy particles that reach the Super-K detector are muons and neutrinos.  The muon flux is $6.0 \times 10^5$ m$^{-2}$ hr$^{-1}$ at sea level, and is reduced to 9.6 m$^{-2}$ hr$^{-1}$ at Super-K~\cite{Galbiati2005}, which corresponds to a muon rate in the detector of about 2 Hz~\cite{Hosaka2006}.  It is easy to veto the muons themselves, but they frequently produce relatively long-lived radioactive isotopes through the breakup (``spallation") of stable nuclei directly or, more commonly, through secondary particles produced through muon energy-loss processes.  The spallation rate is large, $\sim 1$ interaction per through-going muon in Super-K, though many of the daughter nuclei are stable or decay in ways that do not produce Cherenkov signals.

Super-K has cuts to reduce backgrounds from the decays of spallation products, but these have to be limited to not overly discard signal events.  Many of the unstable isotopes produced have half-lives of order 1 s, comparable to the time between successive muons.  It is easy to estimate that a simple cut of all events in a cylinder of radius even a few meters around each muon track for a few seconds leads to a detector deadtime of $\sim 20\%$.  The real algorithm used by Super-K is more complex, and is based on a likelihood analysis that takes into account distance and time from the preceding muon as well as a variable related to muon energy loss, but a similar deadtime is achieved~\cite{Koshio, Hosaka2006}.  Even though the Super-K spallation cuts have a rejection efficiency of $\sim 90\%$~\cite{Gando2003}, the remaining background rate is still $\sim 10$ times greater than the solar neutrino signal rate above several MeV (this is then reduced by another factor $\sim$ 10 by the solar direction cut, leaving a background comparable to the signal)~\cite{Abe2011}.  For the DSNB search, a higher energy threshold can be used to dramatically reduce backgrounds, but spallation decays are still overwhelming below about 18 MeV~\cite{Malek2003} (16 MeV with new techniques~\cite{Bays2012}).

Our goal for this paper is to detail the production processes for spallation backgrounds in Super-K and the physical characteristics of where, when, and with what associated particles these decays occur.  With this information, it will be possible to make better cuts to reject backgrounds while preserving signals.  For solar neutrinos, such improvements could help improve the significance of the 2.7-$\sigma$ hint of the day-night effect from neutrino mixing in Earth~\cite{Liu1997,Maris1997,Blennow2004,Renshaw2013}.  They may also help lead to the first detection of the {\it hep} neutrino flux, which is likely only a factor of a few away from detection~\cite{Marcucci2000,Bahcall2001,Fukuda2001,Bahcall2004,Hosaka2006}.  Such measurements would improve our knowledge of the Sun and of neutrino mixing parameters~\cite{Fogli2011,Haxton2012,Bellini2013lSolar}.  Reduction of spallation backgrounds would also help lower the energy threshold in the DSNB search~\cite{Malek2003,Bays2012} to where the signal is larger~\cite{Horiuchi2009,Beacom2010}, which might help lead to a first detection.

Until now, there has been no detailed published study of spallation backgrounds in water.  The Super-K cuts have been developed from empirical studies~\cite{Hosaka2006,Cravens2008,Abe2011, Bays2012}, and not from theoretical calculations.  Further, they treat all isotopes together, without taking into account significant differences in their production, properties, and distributions.  With Super-K nearly reaching the sensitivity needed for the above discoveries, a more detailed approach is needed. The interactions of muons with scintillator have been studied extensively with underground~\cite{Aglietta1989, Hertenberger1995, Boehm2000, Menghetti2005} and accelerator~\cite{Hagner2000,Chazal2002,Lin2013} experiments, and measurements like these have been incorporated into simulation packages like FLUKA~\cite{Ferrari2005,Battistoni2007} and GEANT4~\cite{Agostinelli2002,Allison2006}.  This gives an opportunity to check our work and to understand the expected uncertainties of the simulations.

This paper is not meant to be a comprehensive study.  It is a first step in understanding spallation backgrounds in water-based detectors, beginning with the yields and the average physical distributions of secondaries and isotopes.  In two subsequent papers, we will go further, showing how characteristics of the showers of secondary particles that produce isotopes can be used to tailor better cuts~\cite{Bays2012} and how those would be improved if Super-K gained the ability to detect neutrons by adding dissolved gadolinium~\cite{Beacom2004}. 

This paper is organized as follows.  In Sec.~\ref{sec:setup}, we describe the setup for our simulation.  In Sec.~\ref{sec:energy_loss}, general points about muon energy loss and secondary particle production are discussed. Our main results are in Sec.~\ref{sec:backgrounds}, where we calculate the neutron and isotope yields and study the properties of the induced backgrounds.  Finally, we present our conclusions in Sec.~\ref{sec:conclusion}.

\section{Setup of calculations}\label{sec:setup}

The Monte Carlo code FLUKA (version 2011.2b.3)~\cite{Ferrari2005,Battistoni2007} is used for this work.  It is a comprehensive code for particle energy loss and interactions with matter.  For our purposes, FLUKA simulates all the physics processes relevant for the interactions of muons and their secondaries with water, including electromagnetic processes such as charged-particle ionization and bremsstrahlung, gamma-ray pair production and Compton scattering, and hadronic processes such as pion production and interactions, photo-disintegration, and low-energy neutron interactions with nuclei.  It has been extensively used to simulate muon interactions in underground detectors, e.g., Refs.~\cite{Wang2001,Kudryavtsev2003,Galbiati2005,Mei2006,Abe2010,Bellini2013}.  The FLAIR interface~\cite{Vlachoudis2009} is used when running FLUKA.

Most of the relevant physics processes and libraries are included in the FLUKA defaults.  To make the low-energy neutron treatment more straightforward, the PRECISIOn card was chosen. Some muon processes, such as photo-nuclear and bremsstrahlung, were specifically activated. The new ion transport library was used. 

The first main input for our simulation is the detector setup.  The Super-K detector is a cylinder of water of diameter 39.3 m and height 41.4 m~\cite{Hosaka2006}.  The outer detector (OD) is separated from the inner detector (ID) by a layer of photomultiplier tubes, most inward-facing, some outward-facing.  The ID is about 2.5 m away from the edge of the detector~\cite{Fukuda2003,Abe2014}.  Our results are calculated only in the fiducial volume (FV) region, which is a virtual cylinder with each side 2 m away from the ID (and about 4.5 m from the outer edge of the OD), containing 22.5 kton of water~\cite{Hosaka2006}.  Water is one of the FLUKA pre-defined materials, including the natural abundances of hydrogen and oxygen isotopes.  Muons may also interact with the surrounding rock to produce showers that enter the detector and produce isotopes.  In the geometry setup, we include 2 m of rock outside the detector to induce secondary production (see Refs.~\cite{Zbiri2010,Empl2012}), though it has only a modest effect.

\begin{figure}[t]
    \begin{center}                  
        \includegraphics[width=\columnwidth]{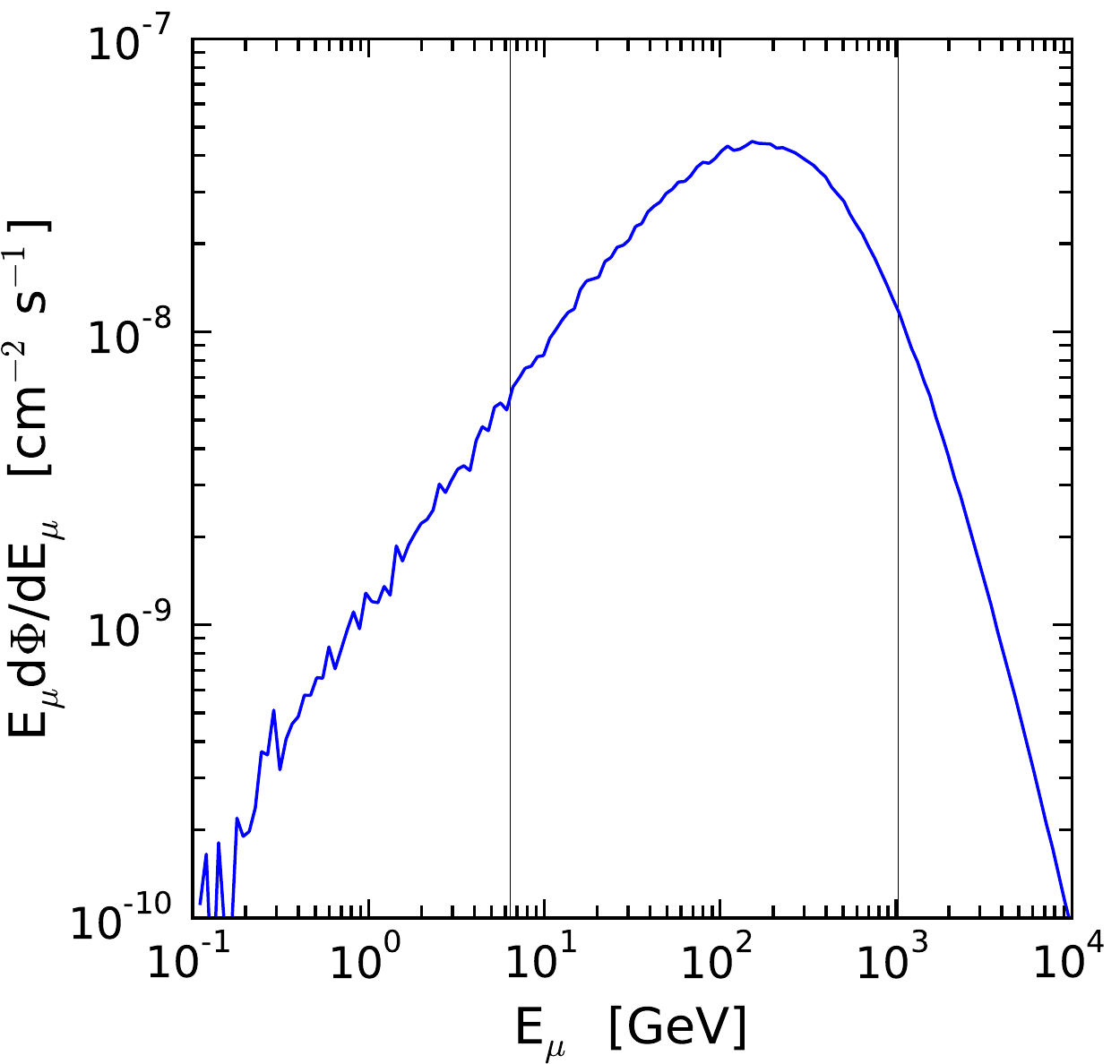}
        \caption{Simulated cosmic-ray muon flux spectrum (integrated over angles) at Super-K~\cite{Tang2006}.  The line near 6 GeV is the minimum ionization energy loss for a vertical muon passing through the Super-K FV.  The line near 1000 GeV is the muon critical energy, above which radiative energy losses dominate.  The fluctuations are from limited statistics in the simulation and are not significant.}
        \label{muon_spectrum}
    \end{center}
\end{figure}

The other main input for our simulation is the muon energy spectrum shown in Fig.~\ref{muon_spectrum}.  The curve is the simulated muon flux at Super-K~\cite{Tang2006}.  Because the muon energy is plotted on a log scale, the flux is plotted as Ed$\Phi$/dE $=$ 2.3$^{-1}$d$\Phi$/dlog$_{10}$E, so that the integrated number of particles per decade (or other interval of fixed multiplicative width) is proportional to the value of this curve (i.e., plotting just d$\Phi$/dE underweights the importance of high-energy bins).  The two vertical lines indicate characteristic energies.  The one near 6 GeV is the minimum ionization energy loss for muons passing vertically through Super-K.  Muons with less energy stop in the detector (as shown in the figure, these are only $\sim$ 5\% of all muons).  The line near 1000 GeV is the muon critical energy, at which the radiative energy loss equals the ionization energy loss.  Muons with higher energies are more likely to produce showers, and thus more isotopes.

By number, most muons are in the range 30 -- 700 GeV, with an average energy of 271 GeV~\cite{Tang2006}.  The spectrum drops at high energies due to the falling spectrum of cosmic rays and at low energies due to muon energy loss in the rock above Super-K.  Integration of the spectrum gives a muon rate at Super-K of 1.8 Hz~\cite{Tang2006}, which is consistent with the published values of 2 -- 3 Hz~\cite{Koshio,Fukuda1999,Habig2001,Gando2003}.  Specifying the muon rate more precisely requires knowing unpublished details about the muon multiplicity, path length and angular distributions, and stopping fraction. Other studies have shown that the detailed shape of the spectrum, for the same average energy, does not affect the isotope yield much~\cite{Kudryavtsev2003,Abe2010}.

We adopt several simplifications for the primary muons.  All muons in our simulation are vertically down-going.  In reality, most muons are down-going, but not perfectly~\cite{Guillian2007}; Tang {\it et al}.~\cite{Tang2006} show that about 75$\%$ of muons have down-going zenith angle $\cos\theta>0.5$ for KamLAND, which is at the same depth and location as Super-K.  A complete 2D map of the simulated angular distribution of muons at Super-K is given in Ref.~\cite{Tang2006,Guillian2007}.  Muons are sent only along the cylinder center.  These two simplifications do not affect our results.  Super-K has very good reconstruction for muon tracks, and all our secondary and isotope yields are calculated per muon path length.  For muons coming in at an angle or a different spot, it would be easy to rescale our results by the actual muon track length.  Besides single through-going muons, there are also muon bundles and muons that only go though a detector corner.  We focus on single through-going muons, because they are the most common and because the other cases are easily identifiable.  We simulate only $\mu^-$; there are also $\mu^+$, but the isotope yields from $\mu^-$ and $\mu^+$ differ very little~\cite{Abe2010,Bellini2013}, except for nuclear captures of stopping $\mu^-$, which we discuss below.

A similar setup was adopted for the spallation study by KamLAND~\cite{Abe2010}.  In their study, spallation yields were measured experimentally and compared to simulation results from FLUKA.  The Borexino spallation study~\cite{Bellini2013} used both simulation packages FLUKA and GEANT4.  Overall, it was found that there are factor of 2 discrepancies between the calculated yields and also between those and the measured values, which is reasonable, given the hadronic uncertainties and that yields for different isotopes vary by orders of magnitude.

\section{Muon energy loss and secondary production}\label{sec:energy_loss}

The average muon energy loss rate is~\cite{Lipari1991,Dutta2001,Groom2001,Bulmahn2010,pdg}
\begin{equation}
    \frac{dE}{dx} = \alpha(E) + \beta(E) E .
    \label{eq:alpha_betaE}
\end{equation}
The $\alpha$ term corresponds to the continuous energy losses due to the ionization (and excitation) of atomic electrons.  It has a typical value of 2 MeV cm$^2$ g$^{-1}$ and does not change much with muon energy.  The ionization can be separated into a restricted ionization energy loss, which is the ionization with soft collisions and small fluctuations, and delta-ray production, which has hard collisions and large fluctuations~\cite{pdg}.  The $\beta E$ term corresponds to the energy losses due to radiative processes through interactions with atomic nuclei.  For muons at hundreds of GeV, pair production and bremsstrahlung are the most important radiative processes, while photo-nuclear has a small contribution~\cite{Groom2001}.  Pair production is a nearly continuous energy loss, but bremsstrahlung and photo-nuclear energy losses have large fluctuations.  Ionization and radiation losses are equal at about 1000 GeV for muons in water, which defines the muon critical energy $E_c$~\cite{Groom2001}.

\begin{figure}[t]
    \begin{center}                  
        \includegraphics[width=\columnwidth]{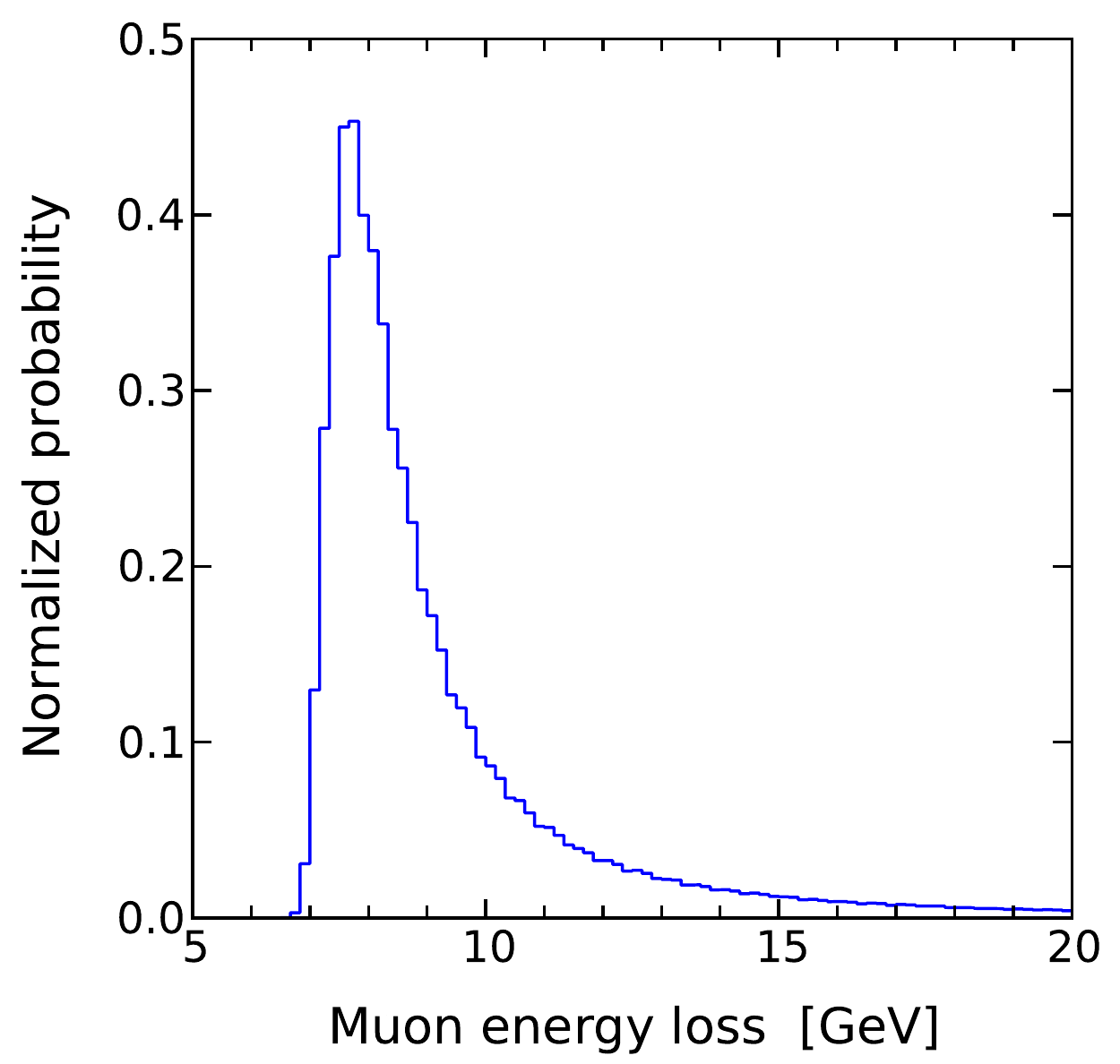}
        \caption{Probability density function of calculated energy loss for vertical through-going muons passing through the Super-K fiducial volume (path length 32.2 m).  The muon energy spectrum used is shown in Fig.~\ref{muon_spectrum}.}
        \label{energy_loss_spectrum}
    \end{center}
\end{figure}

Figure~\ref{energy_loss_spectrum} shows the energy loss distribution for vertical (path length 32.2 m) through-going muons in the Super-K FV.  The restricted ionization energy loss is about 6 GeV and the pair production loss is about 1 GeV.  These two terms have almost no fluctuations and correspond to the minimum energy loss of 7 GeV shown in Fig.~\ref{energy_loss_spectrum}.  On average, muons lose about 11 GeV, which means 4 GeV for the total of the delta-ray production, bremsstrahlung, and photo-nuclear processes.  Bremsstrahlung energy loss is primarily responsible for producing the high energy loss tail~\cite{pdg}. 

Muons lose energy to the production of secondary particles, and there is a lot of energy available to make many of them, as shown in Fig.~\ref{energy_loss_spectrum}.  These interactions do not appreciably affect the parent muon, as the energy loss in the detector is small compared to the muon energy.  The muon interaction cross sections then do not change much as muons lose energy traveling through the detector~\cite{Groom2001}.  The muon tracks have only minor deflections, with 90$\%$ of muons having less than 30 cm transverse displacement when they exit the FV.

\begin{figure}[t]
    \begin{center}                  
        \includegraphics[width=\columnwidth]{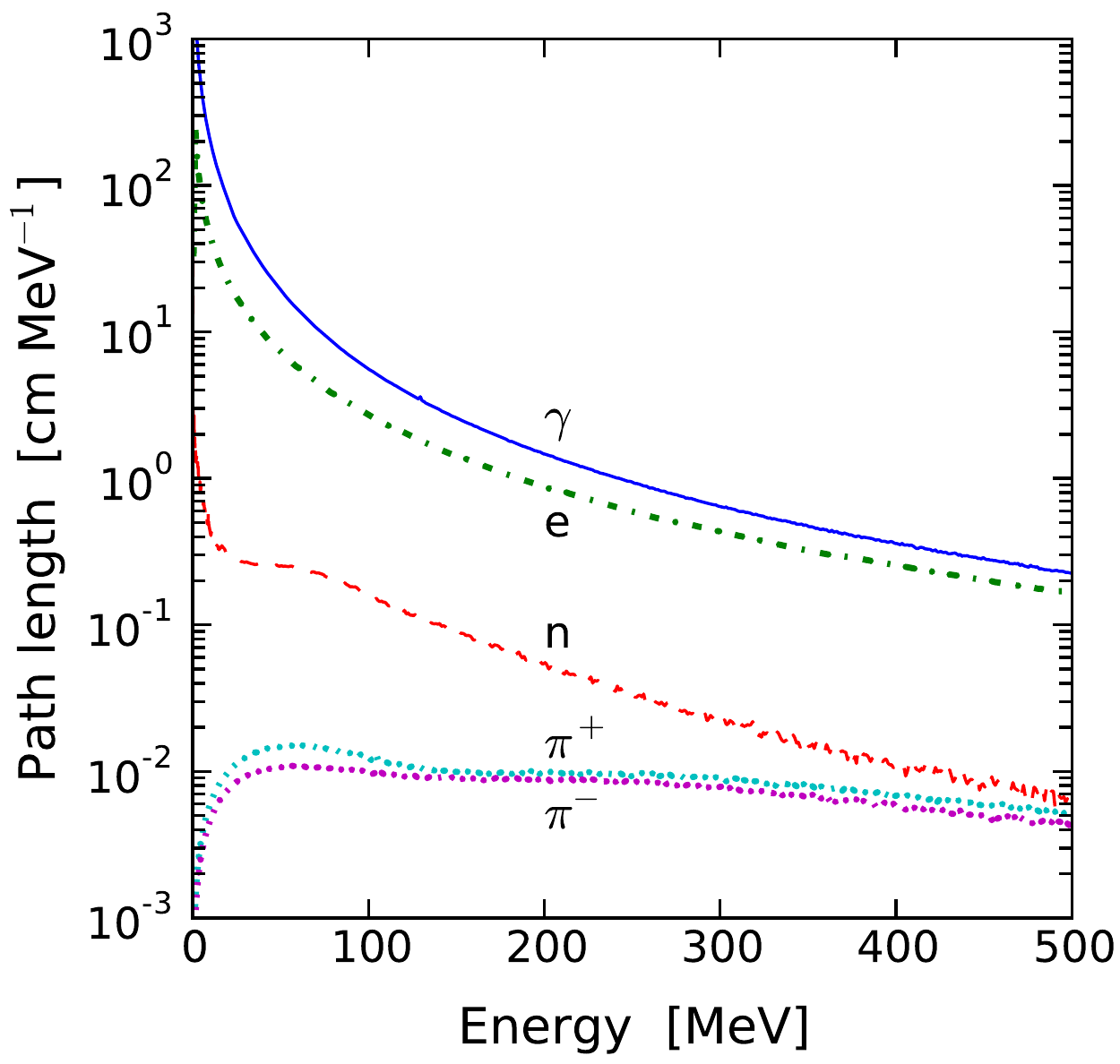}
        \caption{Secondary particle path length spectra made by cosmic-ray muons in Super-K.  The y axis is the cumulative path length, i.e., the total distance traveled by all particles of a given species at each energy, and the x axis is kinetic energy.  Here e means the sum of electrons and positrons.  The proton path length is not shown; it is similar to the pion path length.  The curve for low-energy secondary muons, also not shown, is at or below 10$^{-3}$.  The results are calculated per single muon path length, here the 32.2 m vertical distance in the FV (in contrast, in Table~\ref{tab:isotopes} below, the yields are quoted per cm of muon track, i.e., $\mu^{-1}$ g$^{-1}$ cm$^{2}$).}
        \label{secondary}
    \end{center}
\end{figure}

Figure~\ref{secondary} shows the average production of secondaries by muons in Super-K.  The plotted path length spectrum is the sum of distances traveled by all secondary particles of the same species at certain energy.  It is similar to the particle multiplicity times the mean free path.  The difference is that here a particle contributes to the path length at low energies after it travels some distance at high energies, so there is a pileup of path length from high energy to low energy.  This path length spectrum is the most useful quantity for calculating interactions by these particles.  These results do not depend on density because they are calculated per muon path length (here the vertical distance through the Super-K FV).

As shown in Fig.~\ref{secondary}, the dominant secondaries are gammas, followed by electrons (and positrons).  This makes sense because the primary ways for muons to lose energy other than ionization are delta-ray production, pair production, and bremsstrahlung, all of which are electromagnetic.  In Fig.~\ref{energy_loss_spectrum}, the average radiative muon energy loss is 5 GeV.  The accumulated path length of the secondary electrons and positrons should be $\sim$ 5 GeV / (0.2 GeV/m) $\sim$ 25 m, and the integral of their curve in Fig.~\ref{secondary} is close to this.

A similar figure in Ref.~\cite{Galbiati2005}, which is based on independent calculations, shows secondaries produced by muon interactions in scintillator.  Detailed comparison between Fig.~\ref{secondary} and Ref.~\cite{Galbiati2005} (taking into account the different plotting scales) shows consistent results.  As expected, there is not much difference between muon interactions in water or scintillator for muon energies of hundreds of GeV.  A minor discrepancy is that there are more $\pi^+$ than $\pi^-$ in Fig.~\ref{secondary}, whereas it is the opposite in Ref.~\cite{Galbiati2005}.  To check this, we ran a separate simulation without hydrogen and found that the slight difference in our Fig.~\ref{secondary} between $\pi^+$ and $\pi^-$ is due to scattering of $\pi^-$ on free (hydrogen) protons. Our best guess is that the $\pi^+$ and $\pi^-$ curves in the figure of Ref.~\cite{Galbiati2005} are mislabeled.

All of the results presented here are averaged over many muon path lengths.  In fact, secondaries are made primarily in electromagnetic and hadronic showers, not uniformly along muon tracks.  In our simulation runs, we see significant correlated variations in the muon energy loss, secondary production, and isotope production along the muon paths.  This is hinted at by the high particle energies in Fig.~\ref{secondary}.  In our follow-up papers, we will discuss the shower nature of secondary production and how taking it into account can help improve background rejection in Super-K.

Muons interact with oxygen nuclei directly to produce isotopes, but the dominant mechanism to make isotopes is through secondaries breaking up oxygen nuclei.  The most important secondaries in this regard are neutrons, pions, and gammas.  Of all spallation-induced isotopes that cause backgrounds in Super-K, only 11$\%$ are made by muons (7$\%$ are $^{16}$N from stopping muons plus 4$\%$ other isotopes); the rest are made by secondary particles. 

\begin{figure}[t]
    \begin{center}                  
        \includegraphics[width=\columnwidth]{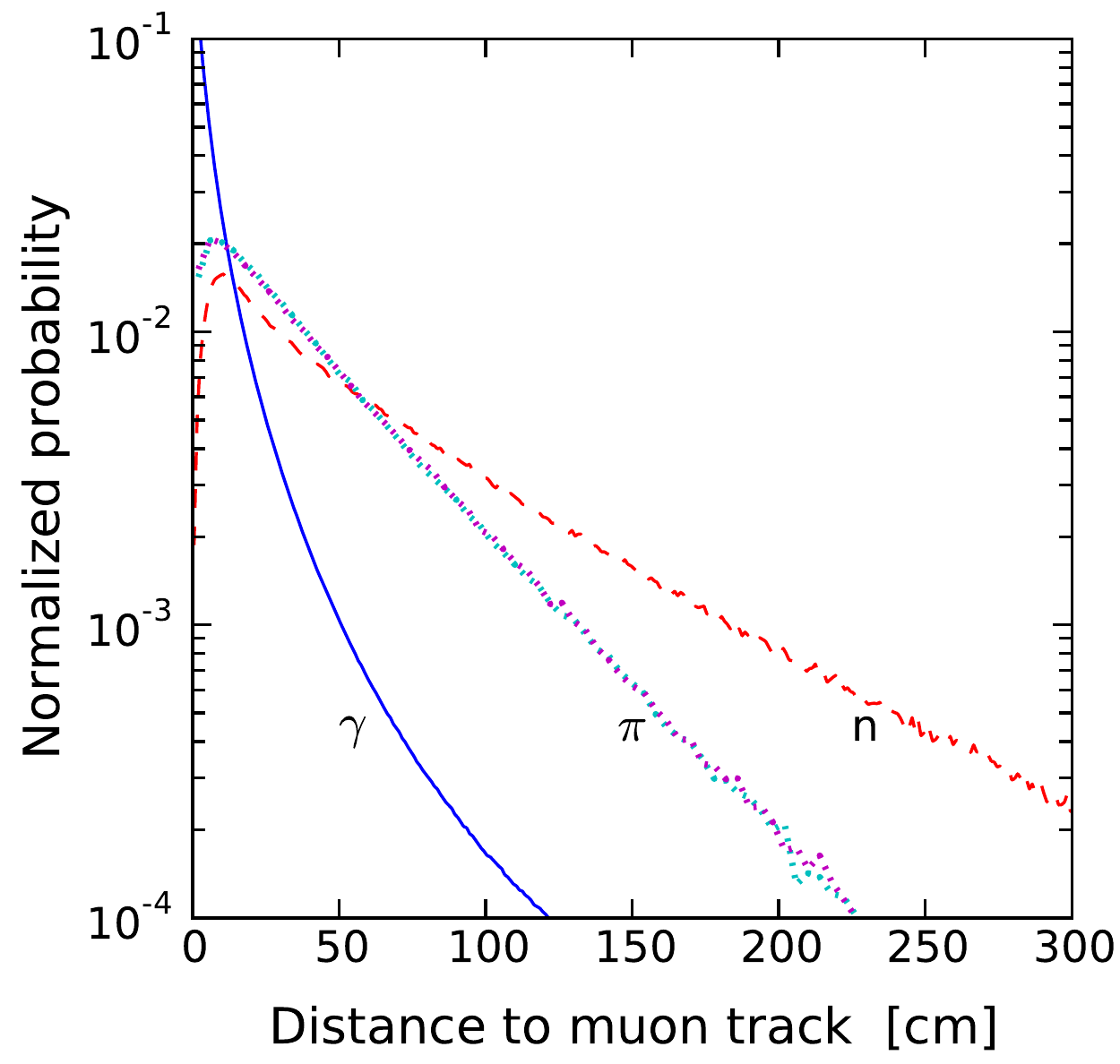}
        \caption{Secondary particle absorption distances to the muon track in Super-K.  Here each distribution is normalized to one.  The plot symbols are the same as in Fig.~\ref{secondary}.  Because of their separate normalizations, the relative heights among the lines should not be compared; e.g., there are not more pions being absorbed at large distances than gammas.}
        \label{secondary_distance}
    \end{center}
\end{figure}

The physical distributions of the secondaries tell us where the isotopes are being made.  The differences reflect how the different secondaries lose energy.  Figure~\ref{secondary_distance} shows the normalized distribution of secondary particle absorption distances to the muon track.  The distribution is dN/dr [cm$^{-1}$], i.e., the area factor $2\pi r$d$r$ is included.  Compared to Fig.~\ref{secondary}, electrons (and protons) are not shown because they are not major parent particles for spallation products.  The gammas have a short mean free path and are mostly forward.  Most gammas are destroyed by pair production, and the Moliere radius (9.8 cm in water~\cite{pdg}) sets a scale for gamma distances from the muon.  The mean free path for pions at these energies is about 1 m~\cite{pdg}.  Assuming pions are destroyed after only one interaction (e.g., $\pi^-$ absorption on p), the falling distribution corresponds to a typical forward direction of $\cos{\theta} \sim 0.9$.  This is consistent with Fig.~\ref{secondary}, where most pions are relativistic.  Among muon secondaries, neutrons travel the furthest from the muon track, with 98$\%$ of neutrons contained within 3 m.  The neutron mean free path is $\sim$ 10 cm above a few MeV, and less at lower energies; neutrons go much farther than this because many scatterings are required to stop them~\cite{Mclane1988}.  The result is very similar to the neutron distance distribution in scintillator~\cite{Bellini2013}.  The carbon number density in scintillator and the oxygen number density in water are comparable, but the cross section for neutrons on oxygen is slightly higher than that on carbon~\cite{Chadwick2011}.  As a result, neutrons travel a bit less far in water.  Compared to the average distance of 74 cm in water, the average distance in scintillator is 81.5 cm~\cite{Bellini2013}.  Most neutrons are absorbed by capture on hydrogen at non-relativistic energies; we also count the reactions of energetic neutrons on oxygen, e.g., (n,p), though this is a small effect.  The Borexino~\cite{Bellini2013} measurement counts only gamma-ray producing captures on hydrogen (mostly) and carbon. 

\section{Isotope and neutron production and distributions}\label{sec:backgrounds}
\begin{table*}[t]
    \begin{center}
    \setlength{\tabcolsep}{4.5pt}
    \setlength{\extrarowheight}{1.5pt}
    \begin{tabular}{ m{1.0em}  d{}   C{3.3cm}   d{}  d{}  C{2.9cm} }
    \hline\hline
        \multicolumn{1}{c}{Isotope} &
        \multicolumn{1}{C{1.8cm}}{Half-life (s)} &
        \multicolumn{1}{c}{Decay mode} &
        \multicolumn{1}{C{2.5cm}}{Yield (total) ($\times 10^{-7} \mu^{-1} \rm{g}^{-1} \rm{cm}^{2}$)} &
        \multicolumn{1}{C{3.5cm}}{Yield (E $>$ 3.5 MeV) ($\times 10^{-7} \mu^{-1} \rm{g}^{-1} \rm{cm}^{2}$)} &
        \multicolumn{1}{c}{Primary process}
        \\ \hline 
        \multicolumn{1}{l}{n} &  &  & 2030 &  &  \\ \hline 
        $^{18}$N & 0.624 &  $\beta^-$   & 0.02  &  0.01 &  $^{18}$O(n,p) \\ 
        $^{17}$N & 4.173 &  $\beta^-n$  & 0.59  &  0.02 &  $^{18}$O(n,n+p) \\  
        $^{16}$N & 7.13 & $\beta^- \gamma$ (66$\%$), $\beta^-$ (28$\%$) & 18 & 18 & (n,p) \\ 
        $^{16}$C & 0.747 & $\beta^-n$ & 0.02 & 0.003 & ($\pi^-$,n+p) \\
        $^{15}$C & 2.449 & $\beta^-\gamma$ (63$\%$), $\beta^-$ (37$\%$) & 0.82 & 0.28 & (n,2p) \\
        $^{14}$B & 0.0138 & $\beta^-\gamma$ & 0.02 & 0.02 & (n,3p) \\
        $^{13}$O & 0.0086 & $\beta^+$ & 0.26 & 0.24 & ($\mu^-$,p+2n+$\mu^-$+$\pi^-$) \\
        $^{13}$B & 0.0174 & $\beta^-$ & 1.9 & 1.6 & ($\pi^{-}$,2p+n) \\ 
        $^{12}$N & 0.0110 & $\beta^+$ & 1.3 & 1.1 & ($\pi^+$,2p+2n) \\
        $^{12}$B & 0.0202 & $\beta^-$ & 12 & 9.8 & (n,$\alpha$+p) \\ 
        $^{12}$Be & 0.0236 & $\beta^-$ & 0.10 & 0.08 & ($\pi^-$,$\alpha$+p+n) \\
        $^{11}$Be & 13.8 & $\beta^-$ (55$\%$), $\beta^-\gamma$ (31$\%$) & 0.81 & 0.54 & (n,$\alpha$+2p) \\
        $^{11}$Li & 0.0085 & $\beta^-n$ & 0.01 & 0.01 & ($\pi^+$,5p+$\pi^+$+$\pi^0$) \\
        $^{9}$C & 0.127 & $\beta^+$ & 0.89 & 0.69 & (n,$\alpha$+4n) \\
        $^{9}$Li & 0.178 & $\beta^-n$ (51$\%$), $\beta^-$ (49$\%$) & 1.9 & 1.5 & ($\pi^{-}$,$\alpha$+2p+n) \\
        $^{8}$B & 0.77 & $\beta^+$ & 5.8 & 5.0 & ($\pi^+$,$\alpha$+2p+2n) \\ 
        $^{8}$Li & 0.838 & $\beta^-$ & 13 & 11 & ($\pi^{-}$,$\alpha$+$^{2}$H+p+n) \\ 
        $^{8}$He & 0.119 & $\beta^-\gamma$ (84$\%$), $\beta^-n$ (16$\%$) & 0.23 & 0.16 & ($\pi^-$,$^3$H+4p+n) \\ \hline
        $^{15}$O &    &    & 351 &   &  ($\gamma$,n) \\
        $^{15}$N &    &    & 773 &   & ($\gamma$,p)  \\
        $^{14}$O &    &    & 13  &   & (n,3n) \\
        $^{14}$N &    &    & 295 &   & ($\gamma$,n+p)  \\ 
        $^{14}$C &    &    & 64 &   &  (n,n+2p) \\ 
        $^{13}$N &    &    & 19 &   & ($\gamma$,$^3$H)  \\ 
        $^{13}$C &    &    & 225 &   &  (n,$^2$H+p+n) \\ 
        $^{12}$C &    &    & 792 &   & ($\gamma$,$\alpha$) \\ 
        $^{11}$C &    &    & 105 &   & (n,$\alpha$+2n)  \\ 
        $^{11}$B &    &    & 174 &   & (n,$\alpha$+p+n)  \\ 
        $^{10}$C &   &   & 7.6 &   & (n,$\alpha$+3n) \\ 
        $^{10}$B &   &    & 77 &   &  (n,$\alpha$+p+2n) \\ 
        $^{10}$Be &   &    & 24 &   &  (n,$\alpha$+2p+n) \\ 
        $^{9}$Be &    &    & 38 &   &  (n,2$\alpha$) \\ \hline
        sum &         &    & 3015 & 50 & \\
    \hline\hline
    \end{tabular}
    \caption{Table of isotope yields.  The top part has background isotopes for Super-K.  The bottom part has isotopes that do not cause backgrounds in Super-K, including those that are stable, have long half-lives, or decay invisibly or with a low beta energy.  The yields and production mechanisms are from simulation.  For the 5th column, the Super-K energy resolution has been taken into account in counting events with decay energies above the Super-K analysis threshold of 3.5 MeV, though it makes little difference.  The observed $^{16}$N decay spectrum (including both betas and gammas) is taken from Ref.~\cite{Blaufuss2001}.  For other isotope decays, only beta energies are included (gammas are ignored).  Yields above 100 are rounded off to 3 significant digits; smaller yields are rounded off to 2 significant digits.  Isotopes with yields smaller than $0.01 \times 10^{-7} \mu^{-1}$ g$^{-1}$ cm$^2$ or mass numbers smaller than 8 (all of which are not backgrounds in Super-K) are ignored.}
    \label{tab:isotopes}
    \end{center}    
\end{table*}

Using the muon and secondary data, we calculate the isotope and neutron yields in Super-K using FLUKA. The isotope counts are read from the RESNUCLEi card.  Neutron counts and production channels are taken from a modified mgdraw.f subroutine.  For neutron counts, processes like (n, 2n) are carefully taken into account. 

We began our study by reproducing all of the relevant KamLAND results~\cite{Abe2010}, and extending the isotope yields to include stable isotopes for comparison to the yields of analogous (stable or unstable) nuclei in Super-K.  Consistent results, within a factor of 2, validate our approach.  The results show interesting differences in the physics of spallation in water and scintillator, as discussed in detail below.  

\subsection{Predicted Yields}

Table~\ref{tab:isotopes} shows the neutron and isotope yields per muon along with associated details.  Almost all isotopes made by muons and their secondaries are listed (we skip isotopes with small yields or small mass numbers). Since Super-K can only detect relativistic charged particles, only betas and gammas (through pair production or Compton scattering) can be seen, while decay products such as neutrons, protons, and alpha particles are invisible (neutron captures on protons are very hard to detect~\cite{Zhang2013}).  The top part of the table contains isotopes that $\beta$ decay and thus are backgrounds in Super-K (referred to as background isotopes); the bottom part of the table contains isotopes that are stable, have long half-lives, or decay invisibly.  

The half-lives of the unstable isotopes range greatly, from 0.008 s to 13.8 s.  A timescale to compare to is the average separation between muons, about 0.5 s.  The beta decay spectra are complicated and have various branches.  Here only the dominant decay modes are listed, though our calculations take all modes into account.  Unsurprisingly, many of the spallation isotopes are short-lived and high-energy compared to intrinsic radioactivities.  The half-lives and decay modes are taken from~\cite{Nudat}.  The isotope decay spectra are taken from Ref.~\cite{Blaufuss2001} for $^{16}$N, Ref.~\cite{Winter2006} for $^8$B, and Ref.~\cite{Tuli2013} for all other isotopes. 

The fourth column shows the isotope yields calculated with FLUKA.  These span five orders of magnitude, which is an important point.  As noted, the accuracy of the isotope production rate is only about a factor of 2.  Yet, because the yields among different isotopes are so different, we can still get a good understanding of their relative importance.  Another point is that the production of beta-decaying isotopes is relatively rare.  The sum of unstable isotopes is 58 in the units of the table, corresponding to about 0.02 unstable isotopes per muon (i.e., multiplying by the vertical distance of 3220 cm).  The sum of the stable or invisible isotopes is around 2950, or about 0.9 isotopes per muon. Neutrons are produced with a yield comparable to that of all isotopes.

The current Super-K solar neutrino analysis has a kinetic energy threshold of 3.5 MeV~\cite{Sekiya2013}, and taking this into account changes the importance of different isotopes.  The fifth column shows the production rate of isotopes with decay energy larger than 3.5 MeV.  Of unstable isotopes with high yields, $^{16}$N is cut the least.  For $^{16}$N decay, 66\% of the time there is a 6.1 MeV gamma ray, which leads to an electron-equivalent energy reduced by a factor $\sim$ 1/4~\cite{Blaufuss2001}.  As a result, the beta spectrum is shifted to higher energies, making it unaffected by the 3.5 MeV cut.  The sum of the yields of background isotopes is reduced to 50 in the units of the table.

The last column shows the most important production channel for each isotope.  For most isotopes, there are several production channels, with different parent particles, often of comparable importance.  Statistically, the assignments of parent particles in the simulation are correct.  For low energy neutrons ($E <$ 20 MeV), FLUKA uses a multi-group treatment, so the correlations among daughter particles are not accurate.  In cases where production by neutrons is important, the results provide a good first understanding, but are not accurate descriptions of the actual interactions.

The final states of the production channels for each isotope indicate particles that could possibly be detected in association with creation of the isotope.  (In addition, there will frequently be prompt gamma rays from the de-excitation of daughter nuclei~\cite{Langanke1996,Nussinov2001,Kolbe2002,Ankowski2012}, but the Cherenkov light from their subsequent signals will be buried under that from the muon.)  It may be possible to identify pion decays in some cases.  Protons and alpha particles will almost always be non-relativistic and hence non-detectable.  At present, it is very difficult to detect neutrons in Super-K~\cite{Zhang2013}, though that would change with the addition of gadolinium~\cite{Beacom2004}; neutron captures are prompt (about 200 $\mu$s in pure water and about 10 times shorter if gadolinium is added), so they are efficiently removed by even a short time cut following a muon.  An important application could be identifying the production of $^8$He and $^9$Li, the decays of which can mimic an astrophysical inverse beta signal because there is a beta followed by a neutron capture.  We find that there is frequently a neutron produced in association with these isotopes, so there would be a neutron capture preceding the $^8$He or $^9$Li decay, unlike for a real astrophysical signal event.  However, we caution that further study of the contributing channels is needed.

The production of $^{16}$N was independently calculated in Ref.~\cite{Galbiati2005}.  This is the most abundant background isotope from muons and it has a long half-life.  The dominant way to make $^{16}$N is $^{16}\mathrm{O}(n, p){}^{16}\mathrm{N}$, which has a yield of $14 \times 10^{-7} \mu^{-1}$ g$^{-1}$ cm$^2$, to be compared to the value found by Ref.~\cite{Galbiati2005}, $23 \times 10^{-7} \mu^{-1}$ g$^{-1}$ cm$^2$.  Sudbury Neutrino Observatory has an upper limit on the $^{16}$N yield of 20 -- $25 \times 10^{-7} \mu^{-1}$ g$^{-1}$ cm$^2$~\cite{Aharmim2005}.  All of these are consistent.

In our simulation, we consider only primary $\mu^-$.  Other studies have shown that isotope production by $\mu^+$ and $\mu^-$ typically differs by only a few percent~\cite{Abe2010}.  One exception is stopping $\mu^-$, which can capture on oxygen and make $^{16}$N by $^{16}\mathrm{O}(\mu^-, \nu_\mu){}^{16}\mathrm{N}$.  Stopping $\mu^-$ make $\sim$ 17\% of $^{16}$N, for which Super-K has a separate cut~\cite{Hosaka2006}.  Consequently, if we take primary $\mu^+$ into account, the $^{16}$N yield would change by about 8\%.  For most subsequent calculations and comparisons to Super-K measurements, we ignore the $\mu^+$ correction to isotope production.

\begin{figure*}[t]
    \begin{center}                  
        \includegraphics[width=\columnwidth]{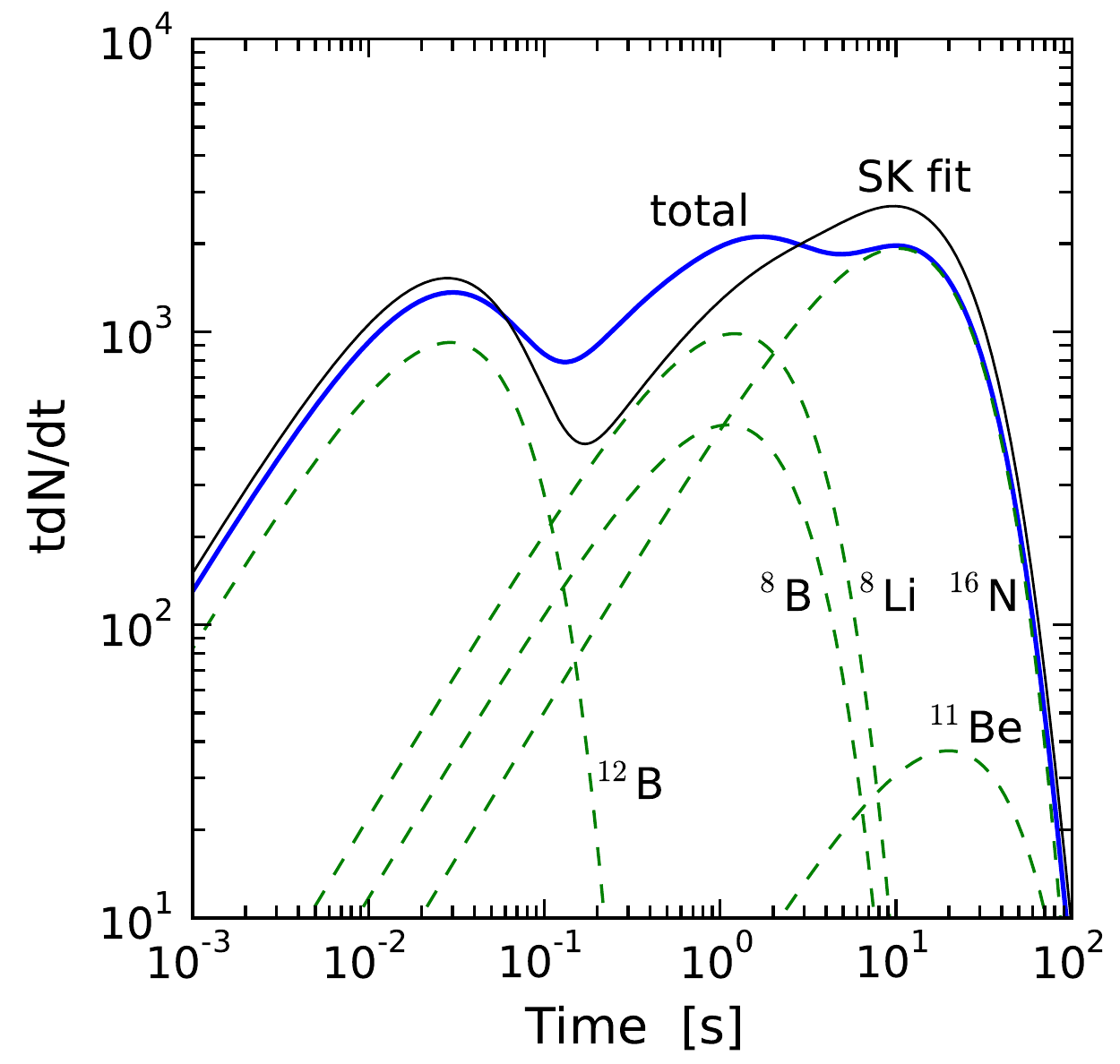}
        \hspace{0.25cm}
        \includegraphics[width=\columnwidth]{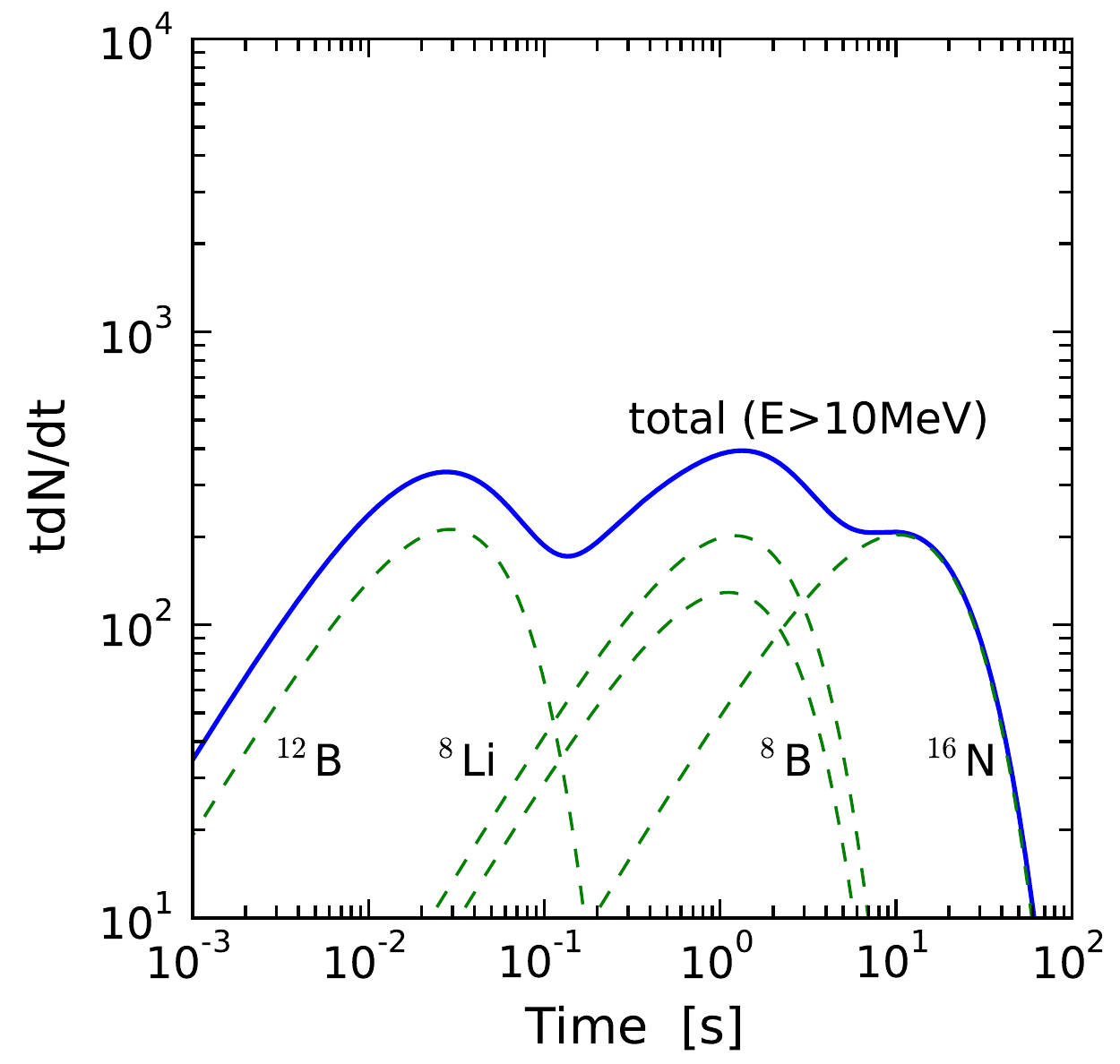}
        \caption{Spallation decay rate distribution.  The y axis is dimensionless, and the relative heights of each curve correctly show their relative contributions.  The $^{16}$N decay spectrum is taken from Ref.~\cite{Blaufuss2001} and the effects of the Super-K energy resolution~\cite{Abe2011} are included.  {\bf Left panel}: The blue line is our FLUKA results, compared to the Super-K empirical fit to spallation-selected data, both with a kinetic energy cut of E $>$ 6 MeV.  The total decay rate is normalized to the Super-K fit, which is measured with high statistics.  The dashed lines show how some example isotopes contribute to the total rate.  {\bf Right panel}: The same, after a 10 MeV kinetic energy cut.}
        \label{decay_time}
    \end{center}
\end{figure*}
 
\subsection{Comparison to Super-K Measurements}

In the following, we focus our comparisons on data above 6 MeV.  At lower energies, detector backgrounds from intrinsic radioactivities are dominant.  The largest intrinsic radioactivity background in the water itself is due to the $^{214}$Bi beta decay following $^{222}$Rn ingress; though its endpoint is 3.26 MeV, energy resolution smears the spectrum to higher energies~\cite{Mitsuda2003,Hosaka2006,Abe2011} (see also Ref.~\cite{Aharmim2013}). There are also radioactivities in the photomultiplier and other detector elements, and these are largely reduced through the fiducial volume cut~\cite{Hosaka2006,Abe2011}.  This dividing line of 6 MeV is in good agreement with the demonstrated effectiveness of the spallation cut above this energy~\cite{Hosaka2006,Cravens2008,Abe2011}, as well as by the results of a dedicated spallation study~\cite{Gando2003}.

Super-K has given a likelihood function of decay time t after the primary muon for decays in a cylinder around the muon path~\cite{Koshio,Hosaka2006}.  This time is well defined because the muon takes only about 100 ns to cross the detector.  The likelihood function is an empirical fit to the sum of all spallation backgrounds, and isotopes with similar half-lives are grouped together.  With the simulated yields from FLUKA, we have each component of this separately.  

Figure~\ref{decay_time} (left panel) shows our combined spallation product decay rate compared to the Super-K fit.  The normalization is chosen so that the integrated event numbers are the same between the simulation and the Super-K fit.  Overall, the total decay rate and the Super-K fit agree well, up to a factor of 2.  The four most abundant isotopes have very different half-lives.  This figure shows how each contributes to the total decay rate on different timescales.  Below about 0.1 s, $^{12}$B is dominant (with a smaller contribution from $^{12}$N, which has a comparable half-life and decay energy); between 0.1 s to 3 s, $^8$Li contributes most; and, after about 3 s, $^{16}$N is dominant.  We also show $^{11}$Be, which has the longest half-life, 13.8 s.  All of the curves in Fig.~\ref{decay_time} (left panel) have a kinetic energy cut of 6 MeV.  

Figure~\ref{decay_time} (right panel) shows a similar result with a 10 MeV kinetic energy cut to the calculation (the similar Super-K measurement is not available).  The main effect of the energy cut is to decrease $^{16}$N compared to other isotopes.  A relatively high energy cut works well for $^{16}$N because of its low endpoint energy.

\begin{figure*}[t]
    \begin{center}                  
        \includegraphics[width=\columnwidth]{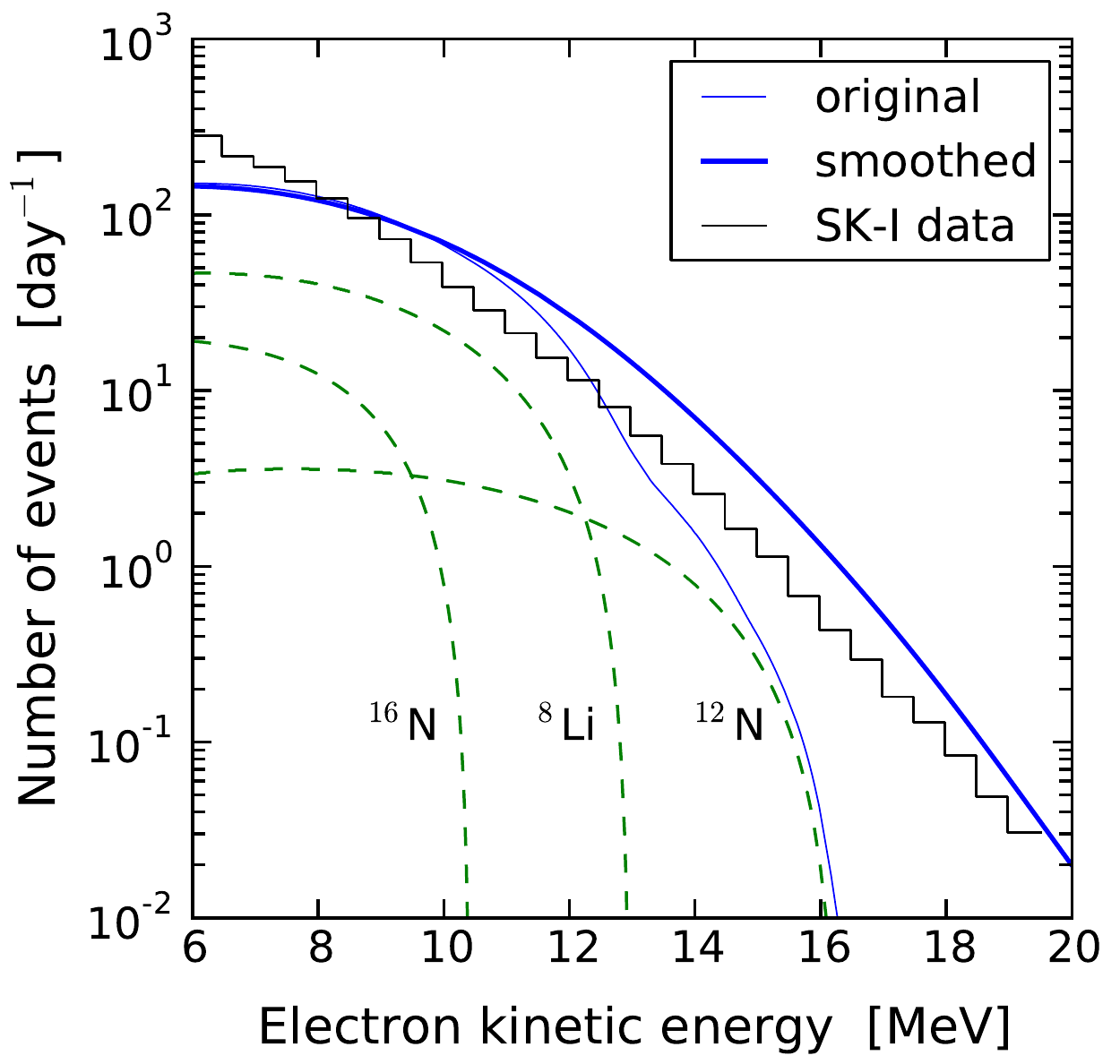}
        \hspace{0.25cm}
        \includegraphics[width=\columnwidth]{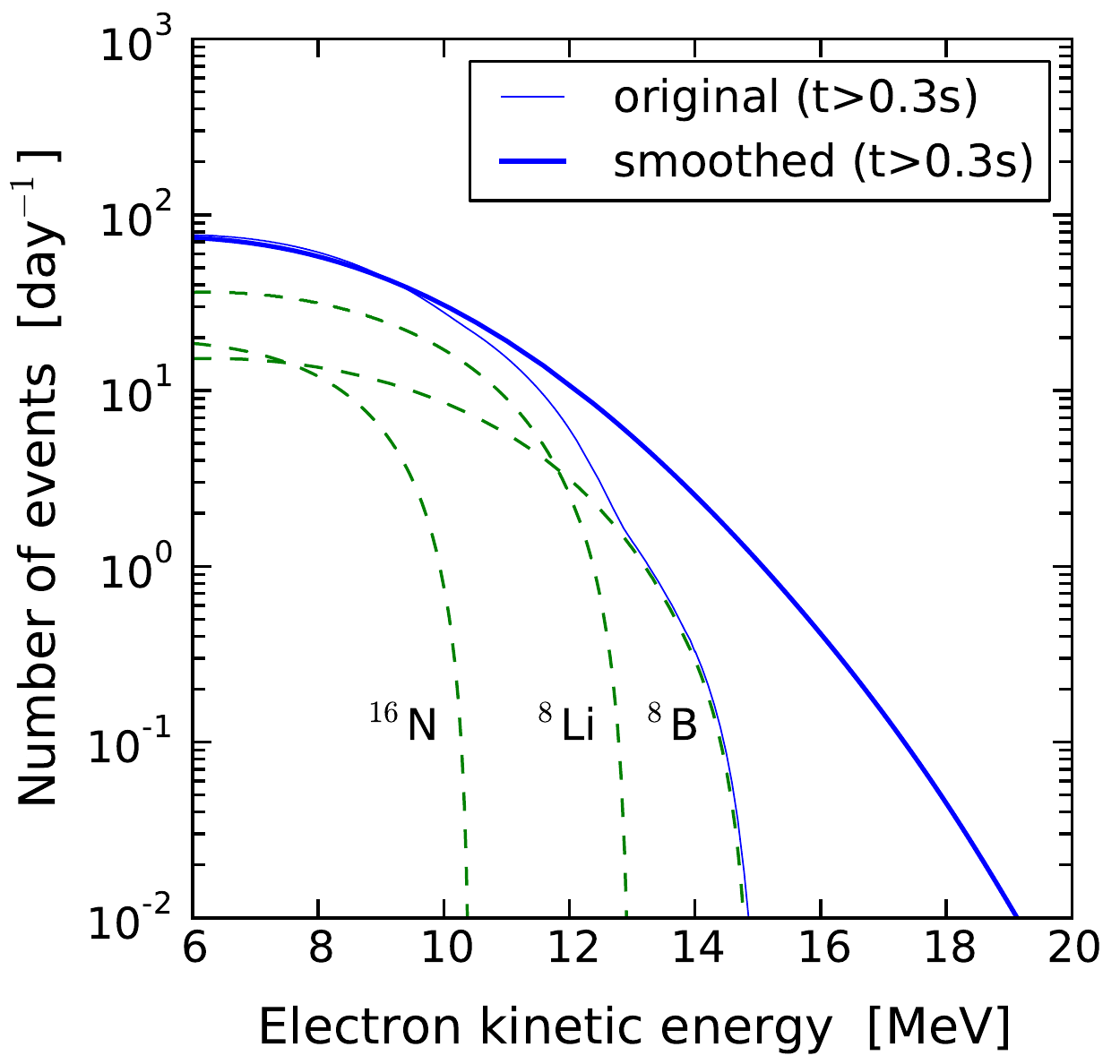}
        \caption{Spallation background energy spectra.  The y axis unit is events per day in the Super-K FV in 0.5 MeV energy bins.  Here the prediction is not normalized to the data.  (In this figure, the expected solar neutrino signal after cuts is $\sim 1$ at low energies, $\sim 0.1$ at medium energies, and vanishing at high energies, as shown in Fig.~39 of Ref.~\cite{Hosaka2006}.)  {\bf Left panel}: The thin blue line shows the total energy spectrum from our FLUKA results, adding up all the component isotope decay spectra, weighted with their yields (shown with dashed lines for some example component isotopes).  The thick blue line is the total spectrum smoothed with the Super-K energy resolution~\cite{Abe2011}; the component spectra are shown before smoothing.  The black stepped line shows the Super-K measurement of the total background spectrum before spallation cuts~\cite{Gando2003}, which is measured with high statistics.  For normalization, the Super-K FV muon rate of 1.88 Hz and the mean muon path length of 32.2 m are used.  Gamma energies are not included in these spectra, as doing so would have only a small effect (it would matter most for $^{16}$N, but that is a subdominant component here).  {\bf Right panel}: The same, after a 0.3 s time cut.}
        \label{gando_spec}
    \end{center}
\end{figure*}

Another comparison we can make with Super-K results is the energy spectrum of spallation backgrounds in the FV.  Similar to above, Super-K has the total decay energy spectrum from all background isotopes~\cite{Gando2003}.  With the simulated yields, adding up the component spectra from all isotopes gives a total spectrum that can be compared to data.

Figure~\ref{gando_spec} (left panel) shows that the simulation and the measurement agree quite well above 6 MeV.  For this comparison, the isotope yields were multiplied by the average muon rate at Super-K (1.88 Hz) and the average muon track length in the FV (32.2 m).  Both numbers have uncertainties because we do not know the precise definitions used by Super-K.  This, together with the limitations of the simulation, introduce the biggest uncertainties.  Taking energy resolution into account is important: the high energy events seen in the detector are mainly from imperfectly reconstructed lower energy events.  The agreement validates our results, especially because the absolute scale is predicted, not fit.  

Figure~\ref{gando_spec} (right panel) shows the isotope spectra after a t $>$ 0.3 s cut, which is about an order of magnitude less than our estimate of the time needed for a simple cylinder cut around each muon (see Sec.~\ref{sec:intro}).  This is chosen to be short enough to not introduce significant deadtime and long enough to eliminate many short-lived isotopes.  The total spectrum decreases by about a factor of 2.  It also affects the relative contributions of isotopes at different energies.  The dominant component at high energy without a time cut is $^{12}$N; after 0.3 s time cut, it is $^8$B.  The fewer isotopes that contribute, the more effective isotope-specific cuts will be (see below).

The Super-K DSNB analysis of Ref.~\cite{Malek2003} has a lower energy threshold of 18 MeV total energy. The total background rate is $\sim$ 0.2 events per day in the 18 -- 20 MeV energy bin.  The rate in Fig.~\ref{gando_spec} is consistent because the measured data in Ref.~\cite{Malek2003} include an increasing contribution from the decays of invisible muons.

The Super-K $^{16}$N calibration study reports that the production rate of $^{16}$N by stopping muons is 11 per day in an 11.5 kton volume~\cite{Blaufuss2001}.  The rate from our calculation is $3 \times 10^{-7} \mu^{-1}$ g$^{-1}$ cm$^2$.  Taking into account the $\mu^-$ fraction in primary muons and the detector efficiency, we predict 22 events per day.  The origin of the discrepancy is unknown, but the Super-K study reported problems with their measurement~\cite{Blaufuss2001}, so we view this factor of 2 as adequate agreement. 

The fact that our same FLUKA predictions match {\it both} the energy spectrum and the time profile of the Super-K data is a powerful indication that they are accurate.  In the energy spectrum, the components are largely overlapping because of the width of the beta spectra and the effects of energy resolution smearing.  In the time profile, the components are better separated because of the wide range of half-lives.  In combination, these provide strong tests of {\it both} the overall production rate of spallation products and the amplitudes of the many components.

\subsection{Comparison to Yields in Scintillator}

\begin{figure*}[t]
    \begin{center}                  
        \includegraphics[width=\columnwidth]{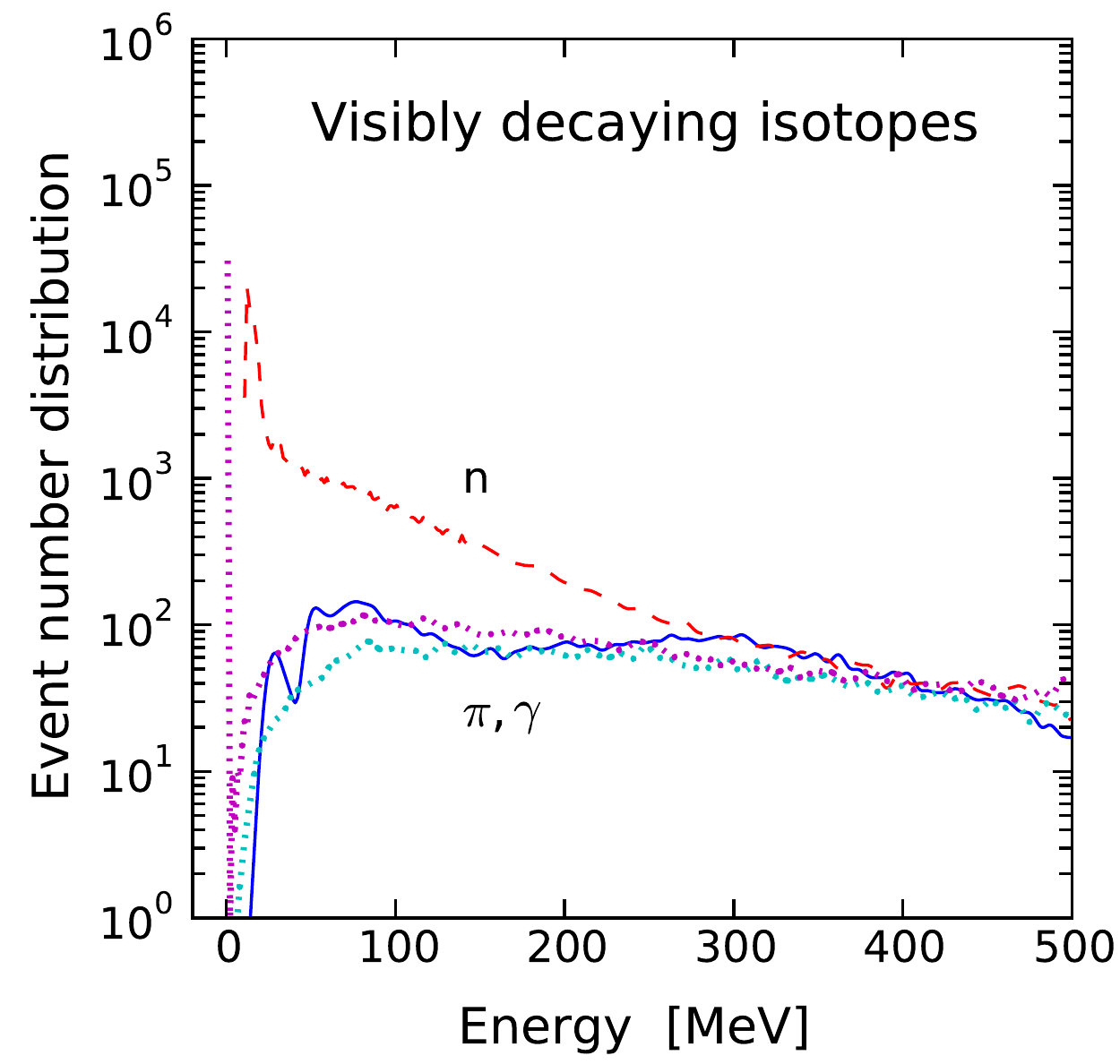}
        \hspace{0.25cm}
        \includegraphics[width=\columnwidth]{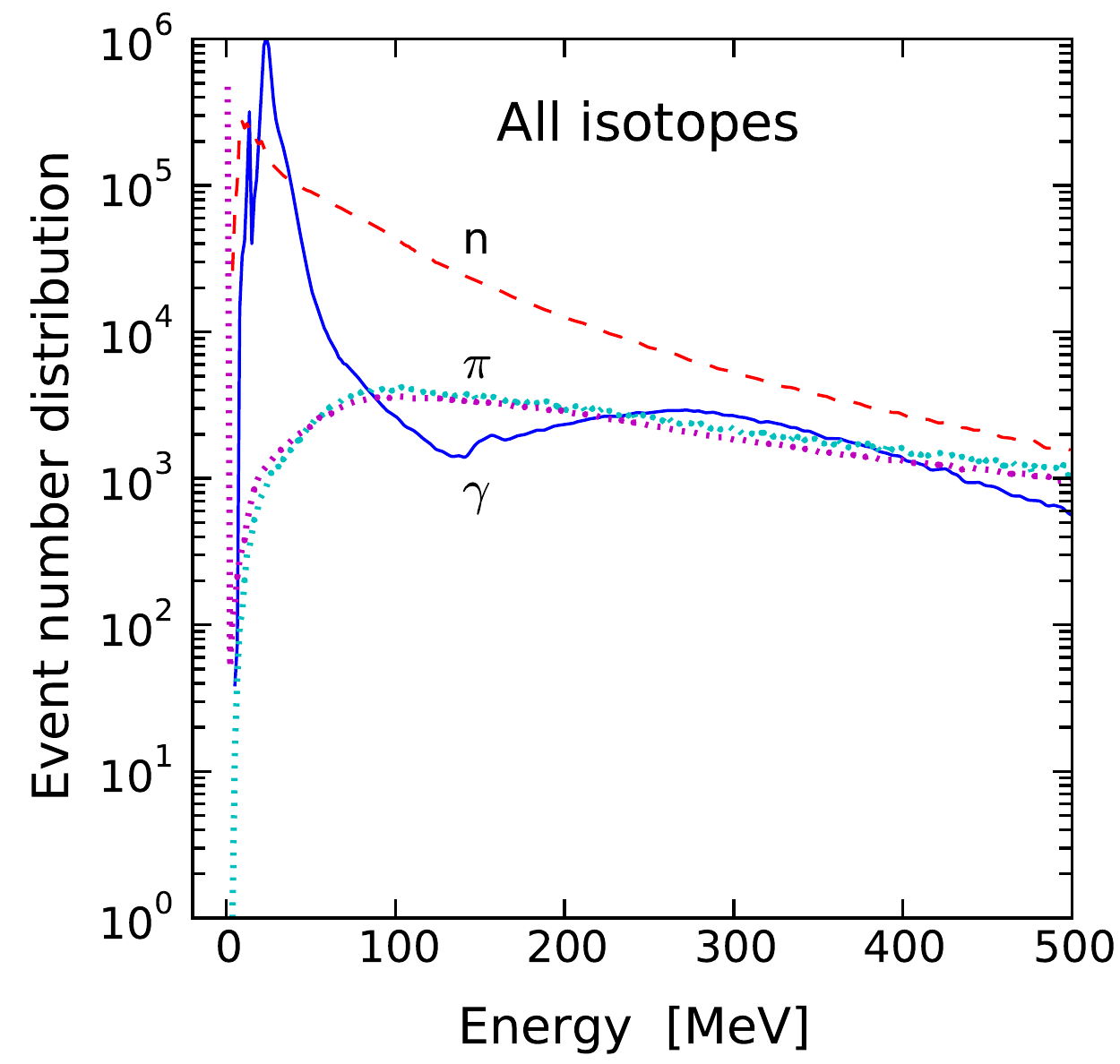}
        \caption{Parent particle kinetic energy spectra in Super-K.  {\bf Left panel}: For background isotopes.  {\bf Right panel}: All isotopes (including the ones that are stable, have long half-lives, or decay invisibly or with a low beta energy).  The absolute normalization for the y-axis is arbitrary (but is the same for both figures); only the relative height and shape matter.  The plot symbols are consistent with those of Fig.~\ref{secondary}.}
        \label{contribution}
    \end{center}
\end{figure*}

A major difference between spallation in scintillator and water is in the absolute background isotope yield.  It is $\sim$ 0.3 of the neutron yield in scintillator, whereas it is only $\sim$ 0.03 of the neutron yield in water.  (In scintillator and water, the neutron yields are similar to each other.)  The reason is that there is a greater fraction of stable or invisibly decaying isotopes produced by muons in water. The neutron number is comparable to the total yield of all isotopes, both in scintillator and water.  It is about 0.7 neutron per muon in the Super-K FV.

The production channels allow us to understand the different spallation processes better.  The isotope yields between scintillator and water are similar if the production mechanisms are similar.  Some of the most abundant isotopes are made by the ($\gamma$,n) reaction, which corresponds to $^{15}$O in water and $^{11}$C in scintillator.  They have yields of 351 and 416~\cite{Abe2010} in the units of $10^{-7} \mu^{-1}$ g$^{-1}$ cm$^2$.  Luckily, $^{15}$O has a low beta-decay energy; in scintillator, $^{11}$C is a serious background.  The most abundant background isotope in water is $^{16}$N, which corresponds to $^{12}$B in scintillator, which has a comparable yield for the same muon path length. 

\subsection{Parent Particle Energy Spectrum}

To understand isotope production mechanisms in more detail, we look at the energy spectra for secondaries making isotopes.  Figure~\ref{contribution} (left panel) shows the spectra of parent particles of spallation background isotopes in Super-K.  Here the y axis is a histogram of event number per MeV with arbitrary absolute normalization.  The relative height reflects how important each parent particle is.  

For making spallation backgrounds in Super-K, the most important parent particle is the neutron, as it contributes almost 10 times more than any others.  The shape of the spectrum is a convolution of the neutron path length shown in Fig.~\ref{secondary} and the neutron-nucleus cross section.  The peak below 20 MeV comes from the $(n,p)$ cross section~\cite{Galbiati2005}.  Due to the nuclear capture of $\pi^{-}$ at rest, there is also a huge peak for low energy $\pi^{-}$.  Gamma, $\pi^+$, and high energy $\pi^-$ contribute roughly equally, each only about half as much as the first $\pi^-$ bin.  The parent particles of fast neutrons are similar to those for isotopes.  Wang {\it et al}.~\cite{Wang2001} showed that at $E_\mu = 270$ GeV, most neutrons are produced by $\pi^-$, followed by gamma and neutron.

One interesting feature is that, even though the dominant secondaries produced directly by muons are gammas and electrons, the ones that make background isotopes in water are mainly hadrons.  This is consistent with the primary processes shown in Table~\ref{tab:isotopes}.  The fact that the gamma and pion curves initially rise with energy is consistent with the path length spectra in Fig.~\ref{secondary}.  The fact that these curves continue to high energies indicates the importance of showers for isotope production.  

As discussed above, the result is somewhat different from muon spallation in scintillator.  A rough count from the KamLAND result tells us that the main parent particle to produce isotopes is gamma, as it is responsible for $^{11}$C and $^7$Be production.  This is consistent with the result shown in Fig.~\ref{contribution} (right panel).  Here we show the parent particle spectra for all isotopes produced in water, including the stable ones and those that decay invisibly.  The gamma contribution is significant, comparable to that of the neutron.  Also, the relative height between the two panels shows the fraction of isotopes that are dangerous in Super-K relative to all isotopes.  The reason for the big difference between the left and right panels is simply that in water, some of the most abundant isotopes made by gammas, e.g., $^{15}$O, $^{15}$N, and $^{12}$C, are invisible in Super-K.  

\subsection{Spatial Distribution of Isotopes}

Because spallation products are produced by muons and their secondary particles, there are spatial and temporal correlations between spallation events and the parent muons.  The muon itself emits Cherenkov light along its entire path, which makes it easy to detect.  Thus, the correlations between muons and isotopes provide an opportunity for physics-motivated cuts.

There are two distances to describe the position of the isotope to the parent muon.  One is the perpendicular distance to muon track, which is one of the variables for the Super-K likelihood function for the spallation cut.  The other is the isotope position along the muon track. 

\begin{figure}[t]
    \begin{center}                  
        \includegraphics[width=\columnwidth]{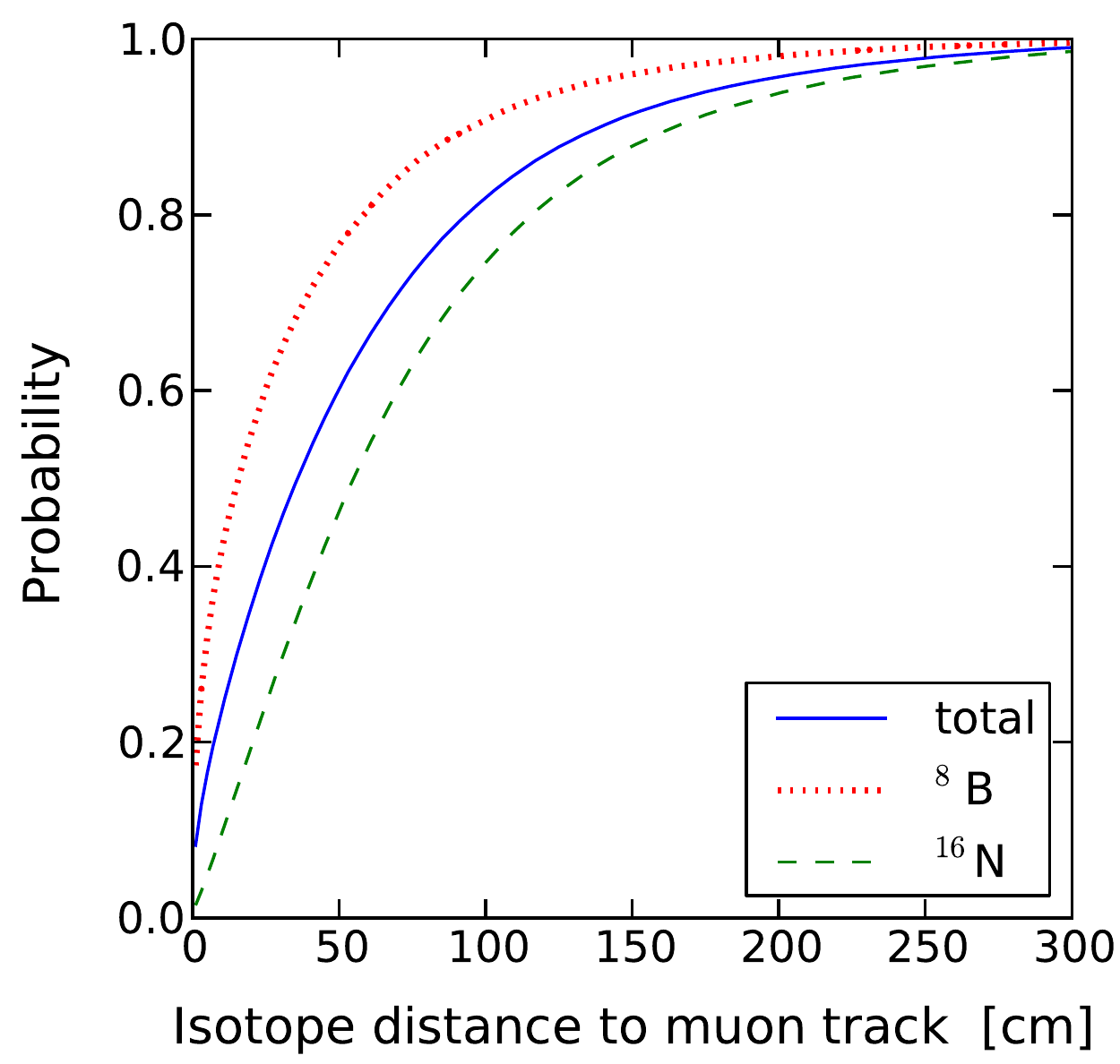}
        \caption{Cumulative distribution of isotope perpendicular distance to the muon track.  The line marked ``total'' is for all isotopes, and the other curves are example isotopes.}
        \label{isotope_distance}
    \end{center}
\end{figure}

Once isotopes are produced, they do not move far before they decay.  Ions stop in a short distance, and there is no significant bulk motion of the water~\cite{Abe2014}.  This can be seen from the fact that the Super-K likelihood function of isotope distance to the muon track shows a peak at very small distance~\cite{Hosaka2006}.  Figure~\ref{isotope_distance} shows our calculated distribution of isotope distance to the muon track.  This shows one of the likelihood functions used for the Super-K spallation cuts.  Our results are consistent with those shown in Ref.~\cite{Koshio} (the Super-K results depend on a variable associated with muon energy loss; we summed over those distributions with appropriate weights).  We did not take the Super-K position resolution into account in Fig.~\ref{isotope_distance}; it is about 1 m at 5 MeV and about 0.5 m at 10 MeV~\cite{Abe2011}.  We find that 99$\%$ of isotopes decay within 3 m.

Each isotope has a different distribution, and we show two examples.  The most abundant background isotope is $^{16}$N, and it dominates the low-energy end of the spectrum.  On the other hand, $^{8}$B contributes the most at the high-energy end, as shown in Fig.~\ref{gando_spec}.  These two isotopes have quite different distributions.  The 90$\%$ containment distance for $^{8}$B is 1.7 times smaller than that for $^{16}$N, which corresponds to a factor of 3 in cylinder volume.  Taking this into account could improve cuts and reduce deadtime.  For example, at high decay energies, $^{8}$B but not $^{16}$N can contribute, so a more specific cut could be used.

Figure~\ref{isotope_distance} shows useful features for improving the Super-K likelihood function for the spallation cut. In the Super-K current likelihood function, the isotope distance to the muon track is one variable.  However, the distance distribution for each isotope can be appreciably different.  As a result, instead of using a combined likelihood function for all background isotopes, a likelihood function for each isotope separately should give a much more accurate description of the physics.

If we consider the isotope distance along the muon track, to first order we would expect a flat distribution when we average over muons (for individual muons, this would have bumps due to showers).  The reason is that, on average, muons have hundreds of GeV energy and lose only about 11 GeV during propagation through Super-K.  More precisely, the isotope yields decrease smoothly from the top of FV to the bottom of FV by several percent.  This is partly due to the decrease of muon energy, and partly due to stopping muons.  There is negligible spillover from the rock above Super-K. 

\section{Conclusions and Future Work}\label{sec:conclusion}

Guided by theoretical understanding and analysis, we use the simulation package FLUKA to study muon interactions with water, the production and properties of secondary particles, and the production and decay of unstable isotopes.  Where possible, we compare our results to published measurements from Super-K, finding good agreement on an absolute scale, i.e., a factor of 2, which is reasonable considering the orders of magnitude differences in production rates.  The residual discrepancies primarily arise from uncertainties in hadronic interactions and unpublished details of the muon backgrounds, and some of the differences could be reduced by calibration to measured data.

As a check, we also performed similar calculations for scintillator-based detectors, for which there are more extensive theoretical studies and experimental measurements.  We focus on comparison to isotope and neutron production in KamLAND~\cite{Abe2010} and Borexino~\cite{Bellini2013}, finding good agreement, within the factors of 2 that have been noted by others between the measurements and calculations and also between calculations with FLUKA versus GEANT4~\cite{Wang2001, Kudryavtsev2003}.  

One interesting point for context is how different the spallation backgrounds are in water Cherenkov detectors compared to scintillator detectors.  First, for water there is the fortunate point that, although the production rate of all isotopes is comparable to that in scintillator, that of unstable isotopes is about ten times less.  Second, many of the unstable isotopes decay without producing Cherenkov light.  Scintillator detectors have the ability to detect neutrons through their radiative captures, and this is a significant advantage in identifying spallation products.  However, if Super-K adds dissolved gadolinium to enable the detection of neutron captures, it will have a similar capability~\cite{Beacom2004}. 

Our calculations for Super-K lead to important new high-level results beyond the details presented here.  First, a demonstration that a theoretical calculation of the spallation backgrounds in water is now possible, even though it was not when Super-K began~\cite{Koshio}.  Compared to an empirical approach, production mechanisms are revealed, aggregates are separated into components, and correlations are preserved.  Second, we show details that were heretofore unavailable.  Important examples are differences between the distributions and correlations of each isotope, including temporal distribution after the muon, distance distribution away from the muon, decay energy spectrum, and associated particles.  

We demonstrate that there is more information to be gained by having likelihood functions of time and distance for each isotope.  Instead of a global likelihood for all spallation decays, our results could be used to construct per-isotope likelihoods that would lead to more precise cuts.  Also, a new variable of decay energy can be used in addition to its original three variables of decay distance to the muon track, decay time, and muon energy loss.  Even modest improvements, say a factor of a few, could lead to significant gains in the ability to measure signals. This could help lead to first discoveries of the day-night effect and the {\it hep} flux in solar neutrinos, as well as the DSNB.

Our results are calculated for Super-K, but they could have wider applicability.  The isotope yields per muon vary only moderately with depth, once that depth is appreciable, because they have a modest dependence on the muon average energy, scaling roughly as $E_\mu^{0.8-1.1}$~\cite{Abe2010}.  As first estimates, our results would provide useful comparisons for the Sudbury Neutrino Observatory~\cite{Aharmim2013}, Hyper-Kamiokande~\cite{Abe2011hk}, and the water shields of a variety of neutrino and dark matter detectors.

It would be valuable for Super-K to produce a dedicated study on spallation backgrounds informed by the predictions of this paper.  The yields of different isotopes could be identified by a global fit that takes into account the full energy and time information on spallation decays, e.g., energy spectra in different time ranges, as has been done for scintillator detectors~\cite{Abe2010,Bellini2013}.  Another key observable is the radial distributions of isotopes produced by different types of secondaries.  An improved FLUKA simulation could be developed using a more complete description of the detector details, especially the muon distributions.  It seems likely that the uncertainties could be reduced to well below a factor of 2 by calibrating the simulation to measured data.

It would also be valuable to have a similar study for the Sudbury Neutrino Observatory~\cite{Aharmim2013}.  The very low muon rate and intrinsic radioactivities would make it easier to identify spallation decays and to avoid confusion over which muon was the parent.  In addition, the ability to detect neutrons would help identify isotope production channels.  With corrections for the different muon spectrum, detector properties, and the production of neutrons by deuterium photo-disintegration, it would be straightforward to relate these measurements to Super-K results.

In two follow-up papers, we will develop further ways to reduce backgrounds in Super-K and other water Cherenkov detectors.  In the first paper, we will study the variations in muon energy loss along the path due to showers, and how this can be used to identify where isotopes are produced.  This effect was discovered empirically in Ref.~\cite{Bays2012}, and our results will provide the first detailed explanation of how it works and how it could be improved.  In the second paper, we will show how the ability to detect neutrons using gadolinium in water, as first suggested in Ref.~\cite{Beacom2004}, can be used to improve cuts to reduce spallation backgrounds.  These papers will include some surprises that will allow significant gains in sensitivity beyond those enabled by results given here.

\section{Acknowledgments}
S.W.L. and J.F.B. were supported by NSF Grant No. PHY-1101216 to JFB.  We thank Sheldon Campbell, Mark Chen, Anton Empl, Cristiano Galbiati, Yusuke Koshio, Pablo Mosteiro, Masayuki Nakahata, Kenny Ng, Itaru Shimizu, Michael Smy, Mark Vagins, and Lindley Winslow for helpful discussions.


\bibliography{yield}
\end{document}